\documentclass[11pt]{article}
\pdfoutput=1

\usepackage{latexsym, graphicx}
\usepackage{amsmath, amssymb, bm}
\usepackage{hyperref}
\usepackage{longtable}
\usepackage{color}

\begin{document}

\begin{center}

\vspace{10mm}

{\LARGE
\textbf{Diffusion in higher dimensional SYK model with complex fermions}
}

\vspace{10mm}

{\large Wenhe Cai$^{\dagger,\ddagger}$\footnote{Email:whcai@shu.edu.cn},~ Xian-Hui Ge$^\dagger$\footnote{Email:gexh@shu.edu.cn},~Guo-Hong Yang$^{\dagger,\ddagger}$\footnote{Email:ghyang@shu.edu.cn}
\\[4mm]
$^\dagger$Department of Physics, Shanghai University, Shanghai 200444, China\\[2mm]
$^\ddagger$Shanghai Key Lab for Astrophysics, 100 Guilin Road, 200234, Shanghai, China
}

\end{center}

\vspace{10mm}

\abstract{We construct a new higher dimensional SYK model with complex fermions on bipartite lattices. As an extension of the original zero-dimensional SYK model, we focus on the one-dimension case, and similar Hamiltonian can be obtained in higher dimensions. This model has a conserved U(1) fermion number $Q$ and a conjugate chemical potential $\mu$. We evaluate the thermal and charge diffusion constants via large q expansion at low temperature limit. The results show that the diffusivity depends on the ratio of free Majorana fermions to Majorana fermions with SYK interactions. The transport properties and the butterfly velocity are accordingly calculated at low temperature. The specific heat and the thermal conductivity are proportional to the temperature. The electrical resistivity also has a linear temperature dependence term.
}

\section{Introduction}
The Sachdev-Ye-Kitaev model is a strongly interacting quantum system at low energy \cite{SY,K}. As the analysis of the recent works \cite{Jevicki:2016bwu,Garcia-Garcia:2016mno,Garcia-Garcia:2017pzl,Bonzom:2017pqs,Gross:2017hcz,Bagrets:2017pwq,Pikulin:2017mhj,Davison:2016ngz}, this model have some interesting properties. It is solvable at large N limit with an emergent conformal symmetry\cite{Maldacena:2016upp}. The reparameterization and gauge symmetries provide connection of SYK to the $AdS_2$ horizon, and the effective action could be obtained from both SYK model and gravity theory on the boundary \cite{Maldacena:2016hyu,Polchinski:2016xgd}. It consists of Majorana fermions with Gauss distribution random at a time. A fermion only moves by entangling with another fermion \cite{SY,Sachdev:2015efa}. The q-body interaction (q is even) is $(i)^{q/2}\sum J_{i_1,...,i_q}\chi_{i_1}...\chi_{i_q}$ with $<J^2_{i_1,...,i_q}>=J^2(q-1)!/N^{q-1}$. This model has maximal chaos \cite{Jensen:2016pah}, and the Lyapunov time $\tau_L=1/\lambda_L$ describes how long a many-body quantum system becomes chaotic. There is a upper bound on the Lyapunov exponent defined in out-of-time correlations in thermal quantum systems\cite{MSS}. The diffusion constants from holography are investigated in previous works, such as \cite{Hartnoll:2014lpa,Fang:2014jka,Ge:2015owa}. Naturally, the transport and diffusivity properties in SYK model could be an interesting aspect. The butterfly velocity is related to the thermal diffusive $D=\upsilon_B^2/\lambda_L$ \cite{Blake:2016wvh,Blake:2016sud}. There are also many new progresses on the generalization of SYK model \cite{Gross:2016kjj,Berkooz:2016cvq,Witten:2016iux,Peng:2017kro,Bonzom:2017pqs} with interesting properties, such as supersymmetry \cite{Fu:2016vas,Peng:2016mxj,Li:2017hdt,Hunter-Jones:2017raw}, chaos \cite{Bhattacharya:2017vaz,Krishnan:2016bvg,Gu:2017ohj}, instability\cite{Bi:2017yvx} and the dual description\cite{Cai:2017nwk}. Several other SYK-like models are studied in \cite{Krishnan:2017ztz}. The higher dimensional of the SYK model is also proposed in \cite{Murugan:2017eto,Narayan:2017qtw}.

On the other hand, there have been various advances in the research on many-body localization transition (MBL) \cite{PH,BSHB,AV,BAA}. The strong-interacting isolated quantum many-body system is localized and fails to approach local thermal equilibrium. Then information about local initial conditions can be locally remembered, and the eigenstates of these systems violate the eigenstate thermalization hypothesis (ETH). There is a transition between the thermal phase in which all the eigenstates satisfy ETH and the many-body localized phase in which all the eigenstates do not satisfy ETH\cite{VHA,PVP,LCR,GADHK,ZSV}. This dynamical transition is an eigenstate phase transition. The validity of ETH in many-body system with local interactions has been proposed in several previous works\cite{Nandkishore:2014kca}, especially SYK models \cite{YLX,HM}. Motived by these facts, we intend to construct a near solvable model(e.g. a generalized SYK model) for exploring the MBL transition.

Recently, Jian and Yao propose a solvable higher-dimensional SYK model exhibiting a dynamical phase transition
between a thermal diffusive metal and an MBL phase \cite{Jian:2017unn}. This 1-dimensional model is defined on bipartite lattices. Each unit cell consists two sites: one site hosting N Majorana fermions with SYK interactions and the other hosting M free Majorana fermions.
Two sublattices are coupled via random hopping. Their calculations show that the dynamic phase transition could be realized by varying the fermion ratio.

In this paper, in order to investigate conductivity and diffusivity properties, we extend the model \cite{Jian:2017unn} to the complex fermion version with a conserved fermion number $Q$ and the chemical potential $\mu$ accordingly. The paper is organized as follows, in section 2, we construct the higher dimensional SYK model with complex fermions and derive the saddle point equations. In section 3, we study the fluctuations of energy and number density. Moreover, we evaluate both thermal and charge diffusion constants. In section 4, we focus on the relationship between diffusion and the butterfly velocity which characterizes quantum chaos. We also investigate transport properties, such as the DC electric conductivity, the thermal conductivity and the heat capacity. The section 5 is the summary and discussion. In the appendix, we give the derivation of the butterfly velocity.

\section{The generalized SYK model}
\begin{figure}[!t]
\centerline{\includegraphics[width=10cm]{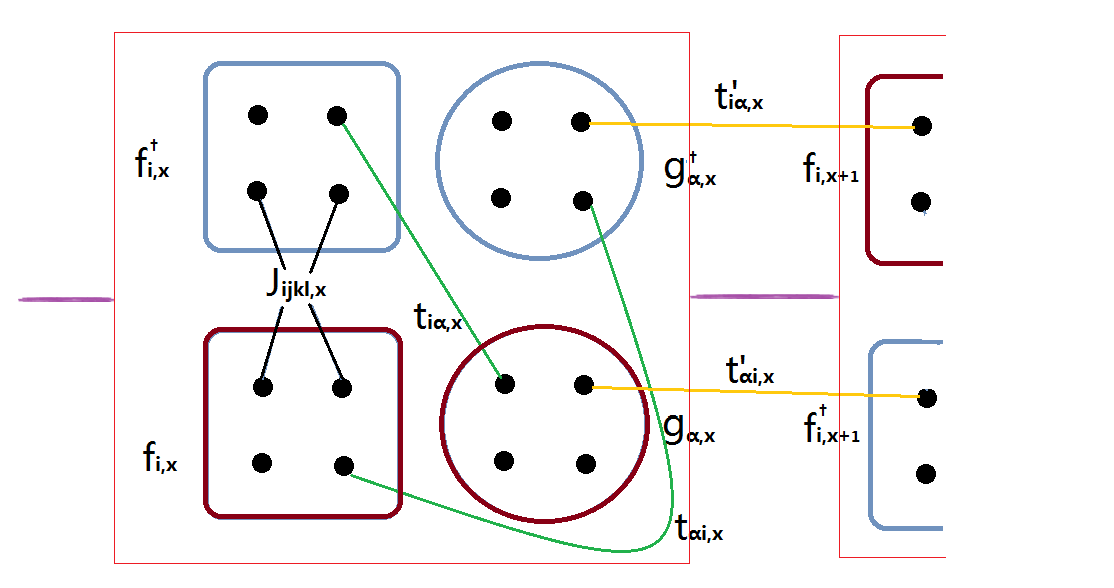}}
\caption{The 1-dimensional model: Each unit cell consists not only two kinds of SYK fermions (creation $f^\dag$, annihilation $f$) on A-sublattice but also two kinds of free fermions (creation $g^\dag$, annihilation $g$) on B-sublattice.}
\label{fg:model}
\end{figure}

Following the approach in \cite{Davison:2016ngz}, we generalize the higher-dimensional SYK model \cite{Jian:2017unn} to the complex fermion version. Unlike the complex SYK model presented in \cite{Gu:2016oyy,SJB} and the model of two $SYK_4$ dots \cite{Chen:2017dav}, this model has N complex fermion with SYK interaction on each site of A-sublattice and M free complex fermions on each site of B-sublattice. Here, the complex fermion can be written in Majorana fermion
\begin{align}
&f_j=\frac{1}{\sqrt{2}}(\chi_{j1}+i\chi_{j2})\,,\,\{f_i,f_j\}=\{f_i^\dag,f_j^\dag\}=0\,,\,\{f_i^\dag,f_j\}=\delta_{ij}\,,\\
&g_\alpha=\frac{1}{\sqrt{2}}(\eta_{\alpha1}+i\eta_{\alpha2})\,,\,\{g_\alpha,g_\beta\}=\{g_\alpha^\dag,g_\beta^\dag\}=0\,,\,\{g_\alpha^\dag,g_\beta\}=\delta_{\alpha\beta}\,.
\end{align}
As illustrated in Figure \ref{fg:model}, our 1-dimensional model comprises of L unit cells, and there are two kinds of sublattices in each unit cell. The coupling $J_{ijkl,x}$ denotes the interaction among fermions on the same sub-lattice, and the coupling $t_{i\alpha,x}$ denotes the interaction among fermions on different sub-lattices. The coupling $t^{\prime}_{i\alpha,x}$ denotes interaction for adjacent cell.

In order to describe a grand-canonical ensemble, we replace the Hamiltonian by
\begin{align}
H&=\sum^L_{x=1}\bigg[\sum_{ijkl}\frac{1}{4!}J_{ijkl,x}f^\dag_{i,x}f^\dag_{j,x}f_{k,x}f_{l,x}+\sum_{i\alpha}(t_{i\alpha,x}if^\dag_{i,x}g_{\alpha,x}+t^{\prime}_{i\alpha,x}ig^\dag_{\alpha,x}f_{i,x+1})\bigg]\nonumber\\
 &-\mu_f\sum_i f^\dag_{i,x}f_{i,x}-\mu_g\sum_i g^\dag_{i,x}g_{i,x}+H.c.
\end{align}
where x labels the lattice site, L is the length of the chain. Note that another choice for the interaction is $(t_{i\alpha,x}if^\dag_{i,x}g_{\alpha,x}+t^{\prime}_{i\alpha,x}ig_{\alpha,x}f^\dag_{i,x+1})$. The coupling $J_{ijkl,x}\,,\,t_{i\alpha,x}\,,\,t^{\prime}_{i\alpha,x}$ are random complex numbers which satisfies $J_{ij,kl}=J^*_{kl,ij}\,,\,t_{i,j}=t^*_{j,i}\,,\,t^{\prime}_{i,j}=t^{\prime *}_{j,i}$ (see \cite{Banerjee:2016ncu,Azeyanagi:2017drg} for details). These Gauss distribution randoms are all anti-symmetric tensor with zero mean obeying
\begin{equation}
 <J^2_{ijkl,x}>=3!J^2/N^3\,,\,<t^2_{i\alpha,x}>=t^2/\sqrt{MN}\,,\,<t^{\prime 2}_{i\alpha,x}>=t^{\prime 2}/\sqrt{MN}\,.
\end{equation}
We define the fermion number $\mathcal{Q}_f=\frac{1}{N}\sum_{i,x}<f^\dag_{i,x}f_{i,x}>\,,\,\mathcal{Q}_g=\frac{1}{M}\sum_{i,x}<g^\dag_{i,x}g_{i,x}>$ and the low-energy scaling dimension of the fermion $\Delta_f=\frac{1}{4}\,,\,\Delta_g=\frac{3}{4}$.
We average over disorder by replica trick $\log Z=\lim_{n\rightarrow 0}\frac{Z^n-1}{n}$, and obtain the replica action
\begin{align}
 S&=\sum_{m,x}\int \frac{1}{2}f^{m\dag}_{i,x}(\partial_\tau-\mu_f)f^m_{i,x}+\frac{1}{2}g^{m\dag}_{\alpha,x}(\partial_\tau-\mu_g)g^m_{\alpha,x}\nonumber\\
  &+\sum_{m,m^\prime,x}\int\int \bigg[-\frac{J^2}{8N^3}|f^{m\dag}_{i,x}f^{m^\prime}_{i,x}|^4-(\frac{t^2}{2\sqrt{MN}}|f^{m\dag}_{i,x}f^{m^\prime}_{i,x}||g^{m\dag}_{\alpha,x}g^{m^\prime}_{\alpha,x}|+\frac{t^{\prime 2}}{2\sqrt{MN}}|f^{m\dag}_{i,x+1}f^{m^\prime}_{i,x+1}||g^{m\dag}_{\alpha,x}g^{m^\prime}_{\alpha,x}|)\bigg]\,.
\end{align}
In the large N limit, we introduce the bilocal fields with $O(N)$ symmetry
\begin{equation}
G^{mm^\prime}_f(\tau_1,\tau_2)=\frac{1}{N}\sum^{N}_{i=1}f^m_{i,x}(\tau_1)f^{m^\prime}_{i,x}(\tau_2)\rightarrow \delta^{mm^\prime}G_f(\tau_1, \tau_2)\,.
\end{equation}
Different replica indices $m,m^\prime$ do not interact.

The green function $G_f$ in our model is given as,
\begin{align}
  \delta_{ij}G_f(\tau_1,\tau_2)&=<f^\dag_i(\tau_1)f_j(\tau_2)>\\
                               &=<\chi_{i_1}(\tau_1)\chi_{j_1}(\tau_2)+\chi_{i_2}(\tau_1)\chi_{j_2}(\tau_2)>\nonumber\\
                               &+i<\chi_{i_1}(\tau_1)\chi_{i_2}(\tau_2)-\chi_{i_2}(\tau_1)\chi_{i_1}(\tau_2)>\,,
\end{align}
and $G_g$ is analogous.

After integrating the fermions, the partition function can be written as a path integral with collective modes $G_{f/g}(\tau_1,\tau_2)$ and $\Sigma_{f/g}(\tau_1, \tau_2)$,
\begin{equation}
 Z=\int DG(\tau_1,\tau_2)D\Sigma(\tau_1, \tau_2)exp(-NS)\,,
\end{equation}
with a collective action based on the Luttinger-Ward analysis in \cite{GPS},
\begin{align}
 \frac{S_{col}}{N}&=\sum^L_{x=1}\Big[-\frac{1}{2}[tr\log(\partial_\tau-\Sigma_{f,x}+\mu_f)+rtr\log(\partial_\tau-\Sigma_{g ,x}+\mu_g)]\nonumber\\
 &+\frac{1}{2}\int\int\Big(\Sigma_{f,x}G_{f,x}+r\Sigma_{g,x}G_{g,x}-\frac{J^2}{4}G^4_{f,x}-\sqrt{r}(t^2G_{f,x}G_{g,x}+t^{\prime 2}G_{g,x}G_{f,x+1})\Big)\Big]\,,
\end{align}
$\Sigma$ is the self energy which contains only irreducible graphs. We define $r\equiv M/N$, it changes from Wigner-Dyson distribution ($r<r_c$) to Possion distribution ($r>r_c$). $r_c$ denotes the critical value.

The equation for the self energy in the grand canonical in the large N limit is
\begin{center}
    $\Sigma(\tau)=-(-1)^{q/2}qJ^2[G(\tau)]^{q/2}[G(-\tau)]^{q/2-1}=\left\{
                                                              \begin{array}{ll}
                                                                -4J^2[G(\tau)]^2[G(-\tau)], & \hbox{q=4;} \\
                                                                2J^2[G(\tau)], & \hbox{q=2;} \\
                                                                -iJ^2, & \hbox{q=1.}
                                                              \end{array}
                                                            \right.$
\end{center}
where $\tau=\tau_1-\tau_2$.

In the IR limit, we drop the $\partial_\tau$ term in the effective action, and obtain the self-consistency equations (i.e. the saddle point equation) by taking $\frac{\delta S_{col}}{\delta G_{f/g}}=0,\frac{\delta S_{col}}{\delta \Sigma_{f/g}}=0$ in the large N limit
\begin{align}
  G_{f,x}(i\omega)&=\frac{1}{i\omega+\mu_f-\Sigma_f(i\omega)}\,,\\
  G_{g,x}(i\tau)&=\frac{1}{i\omega_n+\mu_g-\Sigma_g(i\omega)}\,,\\
  \Sigma_{f,x}(\tau)&=-4J^2G^2_{f,x}(\tau)G_{f,x}(-\tau)+\sqrt{r}[2t^2G_{g,x}(\tau)+2t^{\prime 2} G_{g,x-1}(\tau)]\,,\\
  \Sigma_{g,x}(\tau)&=[2t^2G_{f,x}(\tau)+2t^{\prime 2} G_{f,x+1}(\tau)]/\sqrt{r}\,.
\end{align}
The above Schwinger-Dyson equations also can be derived by summing up the one particle irreducible diagrams \cite{Bhattacharya:2017vaz}.

\section{Fluctuations and diffusion constants}
The reparametrization symmetry  maintains the emergent conformal symmetry of the original SYK model, which is spontaneously broken to $SL(2,R)$ leading to zero modes \cite{Maldacena:2016hyu}. $SL(2,R)$ is the isometry group of $AdS_2$. In the complex SYK model, an additional U(1) phase field $\phi$ is needed\cite{Davison:2016ngz}. The conserved U(1) density is related to the fermion number constraint \cite{Sachdev:2015efa} and the chemical potential \cite{Fu:2016yrv}. So under reparametrization of time $\tau\rightarrow h(\tau)$, we have
\begin{equation}
\tilde G_{a,x}(\tau_1,\tau_2)=[h^\prime(\tau_1)h^\prime(\tau_2)]^{\Delta_a}G_{a,x}\big(f(\tau_1),f(\tau_2)\big)e^{i\phi(\tau_1)-i\phi(\tau_2)}\,.
\end{equation}
where $a=f,g$.

If we take $\omega\rightarrow 0$, the saddle point solution is \cite{Bulycheva:2017uqj}
\begin{equation}
  G_f(\tau_1,\tau_2) = \Big(\frac{1-r}{4\pi J^2}\Big)^{1/4}\frac{sgn(\tau_1-\tau_2)}{|\tau_1-\tau_2|^{1/2}}\,.
\end{equation}
Similarly,
\begin{equation}
 G_g(\tau_1,\tau_2)=\frac{1}{2(t^2+t^{\prime 2})}\Big(\frac{r^2J^2}{4\pi^3(1-r)}\Big)^{1/4}\frac{sgn(\tau_1-\tau_2)}{|\tau_1-\tau_2|^{3/2}}\,.
\end{equation}
We redefine $\Sigma(\tau_1,\tau_2)\rightarrow\Sigma(\tau_1,\tau_2)+\delta(\tau_1-\tau_2)\partial_{\tau_2}$ , then the action can be written as $S=S_{UV}+S_{IR}$.

Inspired by the reparametrization symmetry, the effective action around the saddle point is $S_{eff}[h(\tau)]=S_{UV}[\tilde G(h(\tau))]-S_{UV}[G(\tau)]$, where
\begin{align}
  &h(\tau)=\tau+\epsilon(\tau)=\tau+\frac{1}{\sqrt{L}\beta}\sum_{\omega_n,p}\epsilon_{\omega_n,p} e^{-i\omega_n\tau+ipx}\,,\nonumber\\
  &\tilde G_{f,x}=G_f+\delta G_{f,x}\,,\,\tilde G_{g,x}=G_g+\delta G_{g,x}\,,\,\omega_n=\frac{2\pi n}{\beta}\,.\nonumber
\end{align}
Using the formula
\begin{equation}
 \frac{1}{2}\int dt_1\bigg[\partial_1\Big(\frac{\tan(\pi/q)}{2\pi}(1-\frac{2}{q})\Big)^{1/q}{\frac{|f^\prime(t_1)f^\prime(t_2)|}{|f(t_1)-f(t_2)|}}^{2/q}\bigg]_{t_2\rightarrow t_1}=\frac{1}{2\pi J}\int dt_1\{f,t_1\}\,,
\end{equation}
we obtain
\begin{equation}
 \frac{S^{(1)}_{UV,\epsilon}}{N}=\frac{1}{2}\sum^L_{x=1}\Big(\frac{(1-r)^{1/2}}{J}\frac{1}{(4\pi)^{1/4}}+\frac{r^{3/2}J}{(1-r)^{1/2}}\frac{1}{(t^2+t^{\prime 2})}\frac{1}{(4\pi)^{1/4}}\Big)\frac{1}{\pi^{3/4}}\int dt_1 \{f,t_1\}\,,
\end{equation}
and
\begin{equation}
 \frac{S^{(2)}_{UV,\epsilon}}{N}=\frac{1}{2}\sum_x\int\int\sqrt{r}t^{\prime 2}(\delta G_{g,x}\delta G_{f,x}-\delta G_{g,x}\delta G_{f,x+1})\,,
\end{equation}
here
\begin{equation} \{f,t\}=\frac{f^{\prime\prime\prime}(t)}{f^\prime(t)}-\frac{3}{2}\bigg(\frac{f^{\prime\prime}(t)}{f^\prime(t)}\bigg)^2\,.
\end{equation}
$S_{IR}/N$ in saddle point action vanishes for reparametrization modes,
\begin{align}
 \frac{S_{IR}}{N}&=\frac{1}{2}\sum^L_{x=1}-\bigg[tr\log(-\Sigma_{f,x}+\mu_f)+tr\log(-\Sigma_{g,x}+\mu_g)\bigg]\nonumber\\
                 &+\int\int\bigg[\Sigma_{f,x}G_{f,x}+r\Sigma_{g,x}G_{g,x}-\frac{J^2}{4}G^4_{f,x}-\sqrt{r}(t^2+t^{\prime 2})G_{f,x}G_{g,x}\bigg]\,.
\end{align}

The phase field $\phi(\tau)$ is related to the $PSL(2,\mathbb{R})$ transformation $f(\tau)$ as
\begin{equation}
  -i\phi(\tau)=2\pi\mathcal{E} T(\tau-f(\tau))\,,
\end{equation}
$f(\tau)$ belongs to $PSL(2,\mathbb{R})$. The sets of $f(\tau)$ and $\phi(\tau)$ leave the green function and the effective action invariant. Introducing $\tilde \phi(\tau)=\phi(\tau)+i2\pi\mathcal{E} T\epsilon(\tau)$, we propose the effective action $S_{\phi,\epsilon}$ for reparametrization modes $f(\tau_1-\tau_2)$ and $\phi(\tau_1-\tau_2)$ via fluctuation around the saddle-point ($\beta=1/T$),
\begin{align}
 S_{\phi,\epsilon}&=\alpha_3\frac{\pi}{\beta}\sum_{n,p}|\omega_n|(|\omega_n|+D_1p^2)|\tilde\phi(\omega_n,p)|^2\nonumber\\
                  &+\frac{\pi}{\beta}\sum_{n,p}(\alpha_1|\omega_n|+\alpha_2p^2)|\omega_n|\big[\omega_n^2-(\frac{2\pi}{\beta})^2\big]|\epsilon_{\omega_n,p}|^2\,.\label{Sfluc}
\end{align}
where
\begin{equation}
  \alpha_1=\frac{1}{64\pi^2}\bigg(\frac{\sqrt{1-r}}{J}+\frac{J}{t^2+t^{\prime\,2}}\sqrt{\frac{r^3}{1-r}}\bigg)\,,\, \alpha_2=\frac{1}{128\pi}\frac{rt^{\prime\,2}}{t^2+t^{\prime\,2}}\,.
\end{equation}
The first term in (\ref{Sfluc}) is the UV correction for the $\phi$ mode.
Comparing with the generalized effective action proposed in \cite{Davison:2016ngz}
\begin{align}
  \frac{S_{\phi,\epsilon}}{N}&=\frac{K}{2}\int^{1/T}_0 d\tau\bigg(\partial_\tau\phi+i(2\pi\epsilon T)\partial_\tau\epsilon\bigg)^2-\frac{\gamma}{4\pi^2}\int^{1/T}_0 d\tau\{h,\tau\}\label{eq:Seff}\\
                             &=\frac{KT}{2}\sum_{n,p}|\omega_n|(|\omega_n|+D_1p^2)|\tilde{\phi}(\omega_n,p)|^2\nonumber\\
                             &\ +\frac{T\gamma}{8\pi^2}\sum_{n.p}|\omega_n|(|\omega_n|+D_2p^2)(\omega^2_n-4\pi^2T^2)|\epsilon_{\omega_n,p}|^2\,,
\end{align}
we have
\begin{equation}
 \gamma=8\pi^3\alpha_1\,,\,K=2\pi\alpha_3\,,\,\frac{K}{\gamma}=\frac{\alpha_3}{4\pi^2\alpha_1}\,,
\end{equation}
\begin{equation}
  D_2=\frac{\alpha_2}{\alpha_1}=\frac{\pi r\sqrt{1-r}Jt^{\prime 2}}{2[(1-r)(t^2+t^{\prime 2})+r^{3/2}J^2]}\,,\label{eq:D2}
\end{equation}
where $D_2$ is temperature-independent constants which specify diffusivity.

Note that the second part in (\ref{eq:Seff}) is Schwarzian, which corresponds to the action of black hole with $AdS_2\times R^2$ horizon. When $r=1$, $t=0\,,\,t^{\prime}=0$ and the adjacent cell has been decoupled. Since the isolated SYK islands are not connected, the diffusion constant $D_2$ vanishes. When $r\ll 1$, the SYK fermions dominate such that a finite thermal diffusion constant exist. When $r\gg 1$ the free fermions dominate such that the diffusion constant is imaginary. So there is a phase transition, and $|r-1|$ could be considered as the order parameter.

Since the phenomenological coupling $K,\gamma$ could not be evaluated analytically, we begin the large q expansion expand by small $\Delta=1/q$ at low T with three universal thermodynamics quantities $Q,\mathcal{S},\mathcal{E}$.
The $T\rightarrow 0$ limit of the entropy is given by
\begin{align}
 \mathcal{S}(Q)&=\frac{1}{(M+N)L}(S_f(Q)+S_g(Q))\nonumber\\
               &=\bigg(\frac{1-r}{1+r}\bigg)\bigg[Q\ln\bigg(\frac{1-2Q}{1+2Q}\bigg)+\frac{1}{2}\ln\bigg(\frac{4}{1-4Q^2}\bigg)-\frac{\pi^2}{2}(1-4Q^2)\Delta^2+\mathcal{O}(\Delta^3)\bigg]\,.\label{eq:entropy}
\end{align}
When $Q=0$, (\ref{eq:entropy}) return to Appendix C of \cite{Jian:2017unn}.
To solve the inverse function $2\pi\mathcal{E}=\frac{dS(Q)}{dQ}$, we make an ansatz
\begin{align}
 Q&=Q_1+Q_2\Delta^2+\mathcal{O}(\Delta^3)\nonumber\\ \mathcal{E}&=\mathcal{E}_1+\mathcal{E}_2\Delta^2+\mathcal{O}(\Delta^3)=\frac{1}{2\pi}\ln\bigg(\frac{1-2Q}{1+2Q}\bigg)+2\pi Q\Delta^2+\mathcal{O}(\Delta^3)\,.\nonumber
\end{align}
Taking $\mathcal{E}$ independent, we obtain
\begin{equation}
 Q_1=-\frac{1}{2}\tanh(\pi\mathcal{E})\,,\,Q_2=-\frac{\pi^2}{2}\frac{\sinh(\pi\mathcal{E})}{\cosh^3(\pi\mathcal{E})}\,.
\end{equation}
The grand potential is given by the express \footnote{The free energy (per site and per SYK flavor) with a tunable divergent density of states has been discussed in the recent study \cite{Haldar:2017pyx}.}
\begin{align}
  \Omega&=-T\sum^L_{x=1}\Big[\sum_n tr\ln\big(i\omega_n+\mu_{f/g}-\Sigma(i\omega_n)\big)-\frac{J^2}{2q}\int^{1/T}_0 d\tau G^q_{f,x}(\tau)G^q_{f,x}(1/T-\tau)\nonumber\\
        &-\sqrt{\gamma}\Big(\frac{t^2}{2q}G^q_{f,x}(\tau)G^q_{g,x}(1/T-\tau)+\frac{t^{\prime 2}}{2q}G^q_{f,x}(\tau)G^q_{g,x+1}(1/T-\tau)\Big)\Big]+...
\end{align}
Then, we follow the analysis in \cite{Davison:2016ngz} and obtain
\begin{equation}
  \Omega(\mu,T)=-\bigg(\frac{1-r}{1+r}\bigg)\bigg[T\ln\bigg(2\cosh\frac{\mu}{2T}\bigg)-2\pi\upsilon T\bigg(\cosh\frac{\mu}{2T}\bigg)^{-2}\bigg(\tan\frac{\pi\upsilon}{2}-\frac{\pi\upsilon}{4}\bigg)\Delta^2+\mathcal{O}(\Delta^3)\bigg]\,.
\end{equation}
where $\upsilon$ satisfy
\begin{equation}
 \frac{\pi\upsilon}{\cos(\pi\upsilon/2)}=\frac{\mathcal{J}}{T}\,,\,\mathcal{J}^2=\frac{q^2J^2}{2(2+2\cosh\frac{\mu}{T}))^{q/2-1}}\,.\nonumber
\end{equation}
Considering the thermodynamic relation, we have the free energy, the chemical potential and the entropy
\begin{equation}
  F(Q,T)=\Omega(\mu,T)+\mu Q\,,\,\mu(Q,T)=\big(\frac{\partial F}{\partial Q}\big)_T\,,\,\mathcal{S}(Q,T)=-\big(\frac{\partial F}{\partial T}\big)_Q\,,
\end{equation}
\begin{align}
  K^{-1}&=\big(\frac{\partial \mu}{\partial Q}\big)_T\sim\frac{4T}{1-4Q^2}+(16J-4\pi^2T)\Delta^2+\mathcal{O}(\Delta^3)\,,\\
  \gamma&=-\big(\frac{\partial^2 F}{\partial T^2}\big)_Q\sim[2\pi^2(1-4Q^2)/J]\Delta^2+\mathcal{O}(\Delta^3)\,.
\end{align}
The entropy at low temperature could be obtained as,
\begin{equation}
 \mathcal{S}(Q,T)=\mathcal{S}(Q)-\bigg(\frac{\partial^2 F}{\partial T^2}\bigg)_Q T\,.
\end{equation}
Additionally, we collect the relationship between thermodynamic parameters and diffusion constants according to the effective action derivation directly in Appendix H of \cite{Davison:2016ngz}.
\begin{equation}
 \frac{D_2}{D_1}=\frac{4\pi^2\Delta^2 K}{3\gamma}\,.
\end{equation}
Combined with the $\mathcal{O}(\Delta^2)$ result, we obtain
\begin{equation}
 D_1=\bigg(\frac{1-r}{1+r}\bigg)^2\frac{3\pi Tr\sqrt{1-r}t^{\prime 2}}{(1-r)(t^2+t^{\prime 2})+r^{3/2}J^2}\,.\label{eq:D1}
\end{equation}
When $r\rightarrow 0$, there are almost entirely SYK fermions. As the previous analysis, the system is composed of isolated stacks of SYK Majorana fermions. Therefore, the diffusion constant $D_1$ vanish. When $r\rightarrow 1$, SYK fermions are as many as free fermions. In this case, the diffusion constants also vanish, and the behavior $\varpropto(r_c-r)^{5/2}$ implies a phase transition at $r=r_c=1$.

\section{Transport properties and the butterfly velocity}
In order to investigate the diffusivity and conductivity properties, we characterize transport by two-point correlations of the conserved number density firstly \cite{KM} and obtain transport coefficient by Green-Kubo relation. Similar to the analysis in \cite{Davison:2016ngz,GQS,PSS}, we obtain the dynamic susceptibility matrix
\begin{align}
  \chi_(\kappa,\omega)&=\left(\begin{array}{cc}
                                  <\mathcal{Q};\mathcal{Q}>_{k,\omega} & <E-\mu \mathcal{Q};\mathcal{Q}>_{k,\omega}/T \\
                                  <E-\mu \mathcal{Q};\mathcal{Q}>_{k,\omega} & <E-\mu \mathcal{Q};E-\mu \mathcal{Q}>_{k,\omega}/T \\
                               \end{array}
                            \right)\nonumber\\
                      &=\left(\begin{array}{cc}
                                  K\frac{D_1 k^2}{-i\omega+D_1 k^2} & 2\pi K\mathcal{E}\frac{D_1 k^2}{-i\omega+D_1 k^2} \\
                                  2\pi K\mathcal{E}T\frac{D_1 k^2}{-i\omega+D_1 k^2} & \gamma T\frac{D_2 k^2}{-i\omega+D_2 k^2}+4\pi^2\mathcal{E}^2KT\frac{D_1 k^2}{-i\omega+D_1 k^2} \\
                               \end{array}
                            \right)\nonumber\\
                      &=\big[i\omega(-i\omega+Dk^2)^{-1}+1\big]\chi_s\,.\nonumber
\end{align}
where $\chi_s=\lim_{k\rightarrow 0,\omega\rightarrow 0}\chi(k,\omega)$, $Q\,,\,E$ is the conserved charge and the energy density in per site lattice, which depend on wavevector $k$ and frequency $\omega$
\begin{align}
   D=\left(
           \begin{array}{cc}
               D_1 & 0 \\
               2\pi\mathcal{E}+(D_1-D_2) & D_2 \\
               \end{array}
      \right)\,.
\end{align}
The conductivities matrix is
\begin{align}
     \left(
           \begin{array}{cc}
               \sigma & \alpha \\
               \alpha T & \overline{\kappa} \\
               \end{array}
      \right)
      =D\chi_s=
      \left(
           \begin{array}{cc}
               D_1 K & 2\pi K \mathcal{E} D_1 \\
               2\pi K \mathcal{E} D_1 T & (\gamma D_2+4\pi^2\mathcal{E}^2KD_1)T \\
               \end{array}
      \right)\,,
\end{align}
where $\sigma$ is the DC electric conductivity, $\alpha$ is the thermoelectric conductivity, $\kappa=\overline{\kappa}-(T\alpha^2)/\sigma$ is the thermal conductivity. Then, the Wiedemann-Franz ration could also be given as
\begin{equation}
  L\equiv\lim_{T\rightarrow 0}\frac{\kappa}{T\sigma}=\frac{D_2\gamma}{D_1K}\,.
\end{equation}
From the analysis above, we observe that $D_2$ is really the thermal diffusion constant, as $D_2$ is related to the thermal conductivity. In many cases \cite{Blake:2016wvh,GQS,Patel:2016wdy}, there is a simple relation between the butterfly velocity and the thermal diffusive
\begin{equation}
  D_2=\frac{\upsilon^2_B}{2\pi T}\,,
\end{equation}
and the previous work \cite{Jian:2017unn} show the relationship remains by calculating the growth of out-of-time-ordered four-point function(see Appendix for the relation between the butterfly velocity and the thermal diffusive). We continue to find the relation between the butterfly velocity
and the charge diffusive
\begin{equation}
  D_1=\bigg(\frac{1-r}{1+r}\bigg)^2\frac{3\upsilon^2_B}{\pi J}\,,\label{eq:vD1}
\end{equation}
which depends on the interaction J of our generalized SYK model. Apparently, the relationship between $\upsilon_B$ and the charge diffusion constant, unlike the thermal diffusion constant, does not depend on temperature T. Furthermore, (\ref{eq:vD1}) also shows that the butterfly velocity $\upsilon_B$ vanishes at the critical point $r\rightarrow 1$, which indicates an MBL phase.

Next, we study the conductivity properties at low temperature. By considering the leading and next-to-leading contribution in the large q expansion, we obtain the DC electric conductivity $\sigma$ and the thermal conductivity $\kappa$ in the small T expansion,
\begin{align}
  &\sigma=D_1 K=\bigg[\frac{3\pi}{4}(1-4Q^2)\big(1+\frac{\pi^2}{q^2}(1-4Q^2)\big)-\frac{3\pi J}{q^2 T}(1-4Q^2)^2\bigg]\frac{r(1-r)^{5/2}(1+r)^{-2}t^{\prime 2}}{(1-r)(t^2+t^{\prime 2})+r^{3/2}J^2}\label{eq:dcEC}\,,\\
  &\kappa=\overline{\kappa}-(T\alpha^2)/\sigma=\gamma D_2 T=\frac{\pi^3 T}{q^2}(1-4Q^2)\frac{r\sqrt{1-r}t^{\prime 2}}{(1-r)(t^2+t^{\prime 2})+r^{3/2}J^2}\label{eq:TC}\,.
\end{align}
The first term in the electrical conductivity (\ref{eq:dcEC}) is constant, and the second term in (\ref{eq:dcEC}) is proportional to $1/T$. This result is close to the dependence of the electric resistivity $\rho=1/\sigma$ on the low
temperature in the normal phase of high temperature superconductors \cite{DARSBJBT}. The thermal conductivity $\kappa$ (\ref{eq:TC}) is linear in the temperature . This result coincides with the cuprate strange metal as in\cite{Hartnoll:2015sea,Ge:2016sel}. The approximate behavior in (\ref{eq:dcEC}-\ref{eq:TC}) indicate not only a dynamical phase transition but also an interesting temperature-dependence.
Besides, the heat capacity could be written as
\begin{equation}
  c=\frac{\kappa}{D_2}=\gamma T=\frac{2\pi^2(1-4Q^2)}{J q^2}T\label{eq:capacity}\,,
\end{equation}
which shows that the heat capacity is linearly with the temperature. Because of the experiment on optimally doped YBCO gives $c\sim T$ above the critical temperature \cite{LMWCL}, our result shows a good agreement qualitatively in the normal phase of cuprates.

\section{Conclusion and discussion}
 In this paper, the diffusive property of an extended SYK model with complex fermions is investigated in the large N limit. We derived the collective action and the Schwinger-Dyson equations. With our saddle point solutions in the IR limit, we studied the fluctuations in the effective action. It consists $PSL(2,\mathbb{R})$ and $U(1)$ symmetries. Then, we obtained the quantum transport when the temperature is taken to zero. With the diffusion constants in our model, we calculate the DC electric conductivity, the thermal conductivity,the heat capacity and the relation between the butterfly velocity and the diffusion constants. As the analysis shows in Section 4, our novel results are qualitatively similar to strange metal.

 As noted in (\ref{eq:D2}) and (\ref{eq:D1}), the thermal diffusion constant $D_2$ relates to the coupling $J$, and the charge diffusion constant $D_1$ relates to the temperature $T$. Moreover, we explore the relationship between $\upsilon_B$ and $D_1,D_2$. We notice that the thermal diffusion constant in the complex model is similar to the one in the real model \cite{Jian:2017unn}. When $M=0$ (i.e. $r=0$) which means that there are no free fermions, the relevant components of the Hamiltonian $\sum^L_{x=1}\sum_{i\alpha}(t_{i\alpha,x}if^\dag_{i,x}g_{\alpha,x}+t^{\prime}_{i\alpha,x}ig^\dag_{\alpha,x}f_{i,x+1})-\mu_g\sum_i g^\dag_{i,x}g_{i,x}$ vanishes. Thus, it is natural that the effective action in our model could return to the Gaussian action for the zero-dimensional complex SYK model as presented in \cite{Davison:2016ngz} consistently.

 Besides, as an extension of the original zero-dimensional SYK model, we focus on the one-dimension case. The higher dimension case is straightly replaced with vectors $x\rightarrow \mathbf{x}$. For example, the vectors connecting neighboring unit in the two-dimensional model are $(0,1),(1,0),(1,1)$. However, there are still some subtle uncertainties in our calculations. First, our calculation is based on the large q expand. Therefore, a future analytical study would be interesting and natural. Second, our model is not applied at $r=0,1$. The issue also occurs in the real case as in \cite{Jian:2017unn}, we leave the critical theory for future study.

\section*{Acknowledgements}
  We would like to thank John McGreevy and Shao-Feng Wu for valuable discussions. The study was partially supported by NSFC China (Grant No. 11375110) and Grant No. 14DZ2260700 from Shanghai Key Laboratory of High Temperature Superconductors.

\section*{Appendix: The butterfly velocity and the thermal diffusion constant}
In this Appendix, we verify the relation between the butterfly velocity and the thermal diffusion constant in our model. The connected part of the four-point function is
\begin{align}
  \frac{1}{\sqrt{MN}}F_{fg,xy}(\tau_1,\tau_2,\tau_3,\tau_4)=\frac{1}{MN}\sum_{i,j}&<f^\dag_{i,x}(\tau_1)f_{i,x}(\tau_2)g^\dag_{j,y}(\tau_3)g_{j,y}(\tau_4)>\nonumber\\
                                                                          &-G_f(\tau_1,\tau_2)G_g(\tau_3,\tau_4)\,.\tag{A1}
\end{align}
Using the translation symmetry,
\begin{equation}
  F_{fg,xy}(\tau_1,\tau_2,\tau_3,\tau_4)=\frac{1}{L}\sum_p F_{fg,p}(\tau_1,\tau_2,\tau_3,\tau_4)e^{ip(x-y)}\,,\tag{A2}
\end{equation}
we obtain the out-of-time-ordered correlation function of momentum values
\begin{equation}
  \frac{F_{fg,p}(2\pi+it,\pi+it,\pi/2,3\pi/2)}{G_f(\pi)G_g(-\pi)}\propto\frac{\Delta_f\Delta_g}{\pi^2}\sum_{n\geq2,even}\frac{(-1)^{n/2}n^2\cosh nt}{\alpha_1n^2(n^2-1)+\alpha_2p^2|n|(n^2-1)}\,.\tag{A3}
\end{equation}
According to residue theorem, we extract the exponential growth part with the pole $\alpha_1+\alpha_2 p^2$,
\begin{align}
 \frac{F_{fg,xy}(2\pi+it,\pi+it,\pi/2,3\pi/2)}{G_f(\pi)G_g(-\pi)}
 &\sim -\frac{\Delta_f\Delta_g}{2\pi L}\sum_{p}\frac{\cosh t}{\alpha_1+\alpha_2 p^2}e^{ip|x-y|}\,,\nonumber\\
 \frac{\sum_{i,j}<f^\dag_{i,x}(2\pi+it)f_{i,x}(\pi+it)g^\dag_{j,y}(\pi/2)g_{j,y}(3\pi/2)>}{MNG_f(\pi)G_g(-\pi)}
 &\sim 1-\frac{\Delta_f\Delta_g}{4\pi N\sqrt{\alpha_1\alpha_2}} e^{2\pi T(t-|x-y|/\upsilon_B)}\,,\tag{A4}
\end{align}
leading to the butterfly velocity,
\begin{equation}
  \upsilon_B^2=2\pi TD_2\,.\tag{A5}
\end{equation}

\end{document}